\documentclass[aps,prl,reprint,superscriptaddress]{revtex4-2}

\usepackage{graphicx}
\usepackage{amsmath}
\usepackage{dcolumn}
\usepackage{textcomp}
\usepackage{bm}
\usepackage{color}
\usepackage{float}
\usepackage{amssymb}
\graphicspath{figs/}

\begin{document}

	
	\bibliographystyle{apsrev4-2}
	\title{Extreme ultraviolet laser by single photon process}
	
	
	\author{D. B. Hong}
	\affiliation{Center for Excellence in Superconducting Electronics, State Key Laboratory of Functional Materials for Informatics, Shanghai Institute of Microsystem and Information Technology, Chinese Academy of Science, Shanghai 200050, People’s Republic of China}
	\affiliation{Center of Materials Science and Optoelectronics Engineering, University of Chinese Academy of Sciences, Beijing 100049, People’s Republic of China}
	
	\author{B. K. Xiang}
	\affiliation{State Key Laboratory of Surface Physics and Department of Physics, Fudan University, Shanghai 200438, People’s Republic of China}
	
	\author{T. Wu}
	\affiliation{State Key Laboratory of Surface Physics and Department of Physics, Fudan University, Shanghai 200438, People’s Republic of China}
	
	\author{Z. H. Liu}
	\email{lzh17@mail.sim.ac.cn}
	\affiliation{Center for Excellence in Superconducting Electronics, State Key Laboratory of Functional Materials for Informatics, Shanghai Institute of Microsystem and Information Technology, Chinese Academy of Science, Shanghai 200050, People’s Republic of China}
	\affiliation{Center of Materials Science and Optoelectronics Engineering, University of Chinese Academy of Sciences, Beijing 100049, People’s Republic of China}
	
	\author{Z. S. Tao}
	\email{zhensheng.tao@gmail.com}
	\affiliation{State Key Laboratory of Surface Physics and Department of Physics, Fudan University, Shanghai 200438, People’s Republic of China}
	
	\author{Y. H. Wang}
	\email{wangyhv@fudan.edu.cn}
	\affiliation{State Key Laboratory of Surface Physics and Department of Physics, Fudan University, Shanghai 200438, People’s Republic of China}
	\affiliation{Shanghai Research Center for Quantum Sciences, Shanghai 201315, People’s Republic of China}
	
	\author{S. Qiao}
	\email{qiaoshan@mail.sim.ac.cn}
	\affiliation{Center for Excellence in Superconducting Electronics, State Key Laboratory of Functional Materials for Informatics, Shanghai Institute of Microsystem and Information Technology, Chinese Academy of Science, Shanghai 200050, People’s Republic of China}
	\affiliation{Center of Materials Science and Optoelectronics Engineering, University of Chinese Academy of Sciences, Beijing 100049, People’s Republic of China}
	\affiliation{School of Physical Science and Technology, ShanghaiTech University, Shanghai 201210, People's Republic of China}
	
	\date{\today}
	
	\begin{abstract}
		Generating laser with short wavelength is a bottleneck problem in laser technology. The current applicable table top extreme ultraviolet (EUV) lasers are all generated by multi-photon process, with low efficiency and the record short wavelength of 113.8 nm do not meet some applications. By utilizing metastable helium atoms excited by microwave and irradiated by a resonant laser, here we report the development of a practical 58.4 nm laser by single-photon-excitation related anti-Stokes Raman scattering (ASRS). The conversion efficiency is much higher than that of high harmonic generation (HHG). The same divergence of 1.4 mrad as that of excitation laser indicates its stimulating character. Our results show an applicable path towards up-conversion by single-photon process to generate table top EUV lasers.
	\end{abstract}
	
	
	\maketitle
	\section{1. Introduction}
	Extreme ultraviolet (EUV) sources are widely used in photoelectron spectroscopy\cite{R1,R2,R3} and lithography\cite{R4,R5}. With numerous pressing applications, much effort has been devoted to developing high-performance compact EUV sources. Rare-gas discharge lamps have been used for a long time in laboratory-based photoelectron spectrometers, but the large spatial divergences of such incoherent sources seriously limit their applications. Thus, laser sources are strongly desired to get better resolutions and to study inhomogeneous materials with small domains\cite{R6}. The current available EUV laser with shorter than 113.8 nm wavelength can only be obtained from large accelerators, namely free electron laser, which has limited beam time and accessibility. Because EUV radiation is absorbed strongly in nearly all materials, the currently usable laboratory-based EUV lasers cannot be produced by the conventional resonant cavity and can only be generated by multi-photon process\cite{R7}. 186 nm\cite{R8}, 177.3 nm\cite{R9,R10}, and 113.8 nm\cite{R11} lasers have been developed using the multi-photon processes in Pb vapor, KBe$_2$BO$_3$F$_2$ (KBBF) crystal, and Xe gas respectively, and many excellent works\cite{R12,R13}  have been done with these laser sources to achieve high-resolution unavailable before. However, the record high photon energy of 10.9 eV (113.8 nm) is still not enough to cover the whole Brillouin zone of many materials\cite{R14} and limits its applications. Coherent EUV sources with higher photon energies can be generated by HHG\cite{R15,R16} from near-infra-red lasers. However, a subsequent monochromator is needed to filter unwanted harmonics for photoelectron spectroscopy. Because only short-duration pulses can be applied to yield enough efficiency, they are mainly used for time-resolved photoelectron measurements, and their bandwidths are not narrow enough to achieve energy resolution better than 10 meV due to quantum uncertainty.
Single-photon related ASRS has shown the ability to generate ultraviolet anti-Stokes Raman laser (ASRL)\cite{R17}. ASRL is related to three levels, namely a lower level, a middle metastable storage level, and an upper level. In principle, in this single-photon process, only one photon is needed to pump the electron at the metastable level to the upper level and subsequently decay to the lower level to emit a photon with a shorter wavelength\cite{R18}. Compared with multi-photon process, the efficiency of ASRL can be very high and even achieve unity if the excitation and anti-Stokes lasers are strong enough to couple the levels forming coherent dressed-states and stimulated Raman adiabatic passage (STIRAP)\cite{R19}. 376 nm, 278 nm, and 178 nm ASRLs have been demonstrated in 1982 by creating metastable population inversions through photodissociation of TlCl\cite{R20} and NaI\cite{R21}. Previously, 53.7 nm EUV ASRL was generated using metastable state in helium atoms bombarded by electron beam\cite{R22}. However, due to the low density of metastable atoms, the EUV laser generated by their work with weak intensity and broad divergence cannot be used in practical applications. The challenge of generating EUV ASRL is how to produce high-density atoms at the metastable storage level with energy much higher than that of the lower level.

	\begin{figure*}[htbp]
		\includegraphics[width=\linewidth]{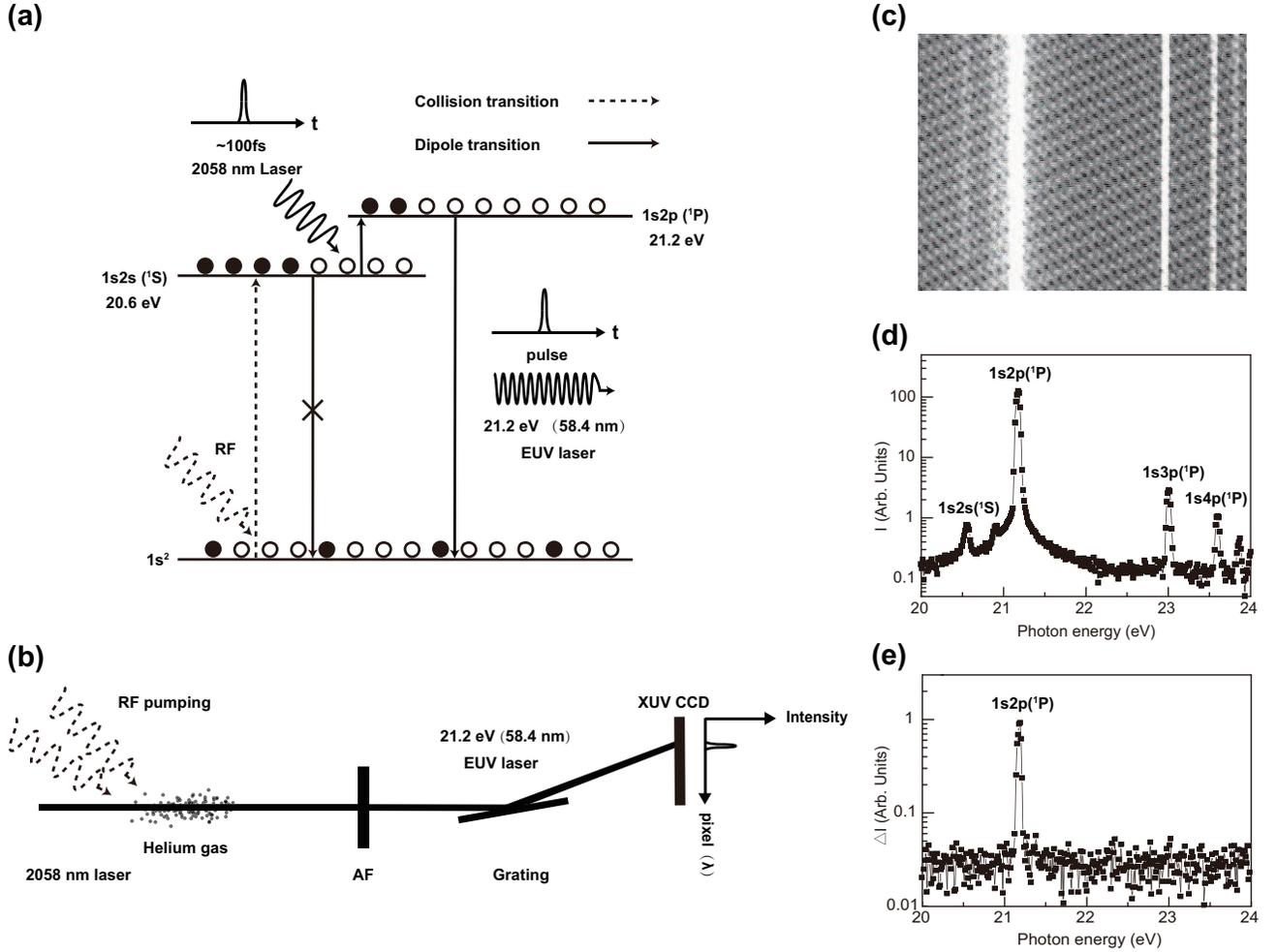}%
		\caption{\label{figs1} Experimental setup and working principle of the EUV laser. (a) Schematic diagram of the generation progress. (b) Schematic diagram of the experimental setup. (c) Grating diffraction spectrum recorded by CCD camera when the helium lamp is turned on and the infra-red laser is turned off. (d) Integrated spectrum corresponding to (c). (e) The difference of spectral intensities with and without the 2058 nm excitation.}
	\end{figure*}

In this work, we obtained high-density metastable helium atoms by microwave excitation, which is a key issue for generating practical EUV ASRL. We successfully developed the 58.4 nm (21.2 eV) EUV laser from the metastable helium atoms by infra-red laser inducing ASRS with a higher conversion efficiency and a simpler structure than those of HHG. The principle to generate EUV laser by single-infra-red-photon excitation is shown in Fig. 1(a). The 1$s$2$s$ ($^1$$S$) atomic level of helium is a metastable state with forbidden dipole transition to the 1$s$$^2$ ground state, so it is possible to excite enough helium atoms to this state by microwave radiation. The atoms in this metastable state can be excited to the 1$s$2$p$ ($^1$$P$) state by the injection of a 2058 nm infra-red excitation laser, and 58.4 nm (21.2 eV) Anti-Stokes photons are subsequently generated by the stimulated decay to the ground state.

	\begin{figure*}[htbp]
		\includegraphics[width=0.5\textwidth,scale=0.5]{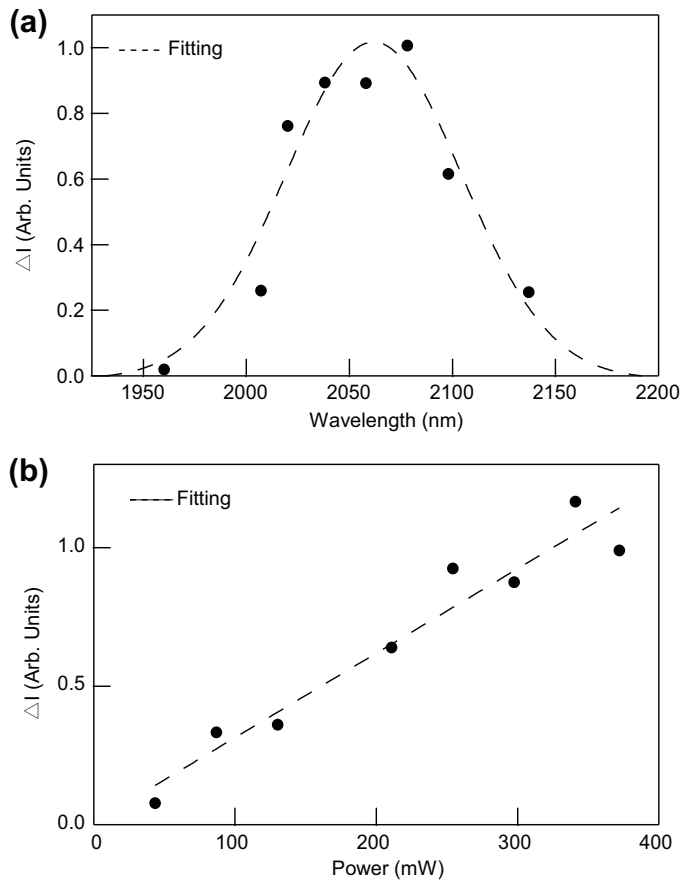}%
		\caption{\label{figs2} The intensities of induced EUV laser. The difference of 58.4 nm peak areas of spectra taken with and without excitation laser on different central wavelengths (a) and powers (b).}
	\end{figure*}

	\section{2. Method and results}
	\section{2.1 Generation of EUV radiation}
	 A schematic diagram of the experimental setup is shown in Fig. 1(b). A commercial helium plasma lamp was used to excite the helium gas by focused microwave. A 6 mm diameter, 70 mm long quartz tube was used to contain the 1 mbar helium gas. The excited helium atoms were created in a 6 mm diameter, 12 mm long cylindrical volume. The infra-red laser with an adjustable central wavelength around 2058 nm, 49.9 nm full width at half maximum (FWHM), and 0.6 W average power is produced by an optical parametric amplifier (OPA) and an 800 nm seed laser with 10 W average power, 100 fs pulse width, and 10 kHz repetition rate. A diffraction grating spectrometer was used to measure the EUV spectrum with a 120-nm-thick aluminum filter (AF) to block the infra-red and visible lights.

	Firstly, we show that the single-photon excitation process can indeed produce EUV photons. Figs. 1(c) and 1(d) show the background spectrum from the helium lamp alone. The peaks at 21.2 eV (58.4 nm), 23.1 eV (53.7 nm) and 23.7 eV (52.2 nm) correspond to the transitions from 1$s$2$p$ ($^1$$P$), 1$s$3$p$ ($^1$$P$) and 1$s$4$p$ ($^1$$P$) levels to the ground state, respectively. Fig. 1(e) shows the difference spectrum by subtracting the background from that obtained with infra-red laser excitation. In contrast to the background emission, the laser-induced EUV photons only exhibit a single peak at 21.2 eV (58.4 nm), which strongly suggests that the infra-red laser drives the single-photon excited ‘$\Lambda$’ process as expected.

	\section{2.2 Resonance of single-photon process}
	To further verify the single-photon mechanism shown in Fig. 1(a), we swept the central wavelengths of the infra-red laser. Fig. 2(a) shows the intensities of induced 58.4 nm photons normalized by the infra-red laser powers at different central wavelengths. The peak around 2058 nm indicates that the induced photon is caused by the excitation from 1$s$2$s$ to 1$s$2$p$ level. This result and the linear dependence of EUV intensity and the power of 2058 nm laser shown in Fig. 2(b) confirm the single-photon mechanism.
	
	\begin{figure*}[htbp]
		\includegraphics[width=\linewidth]{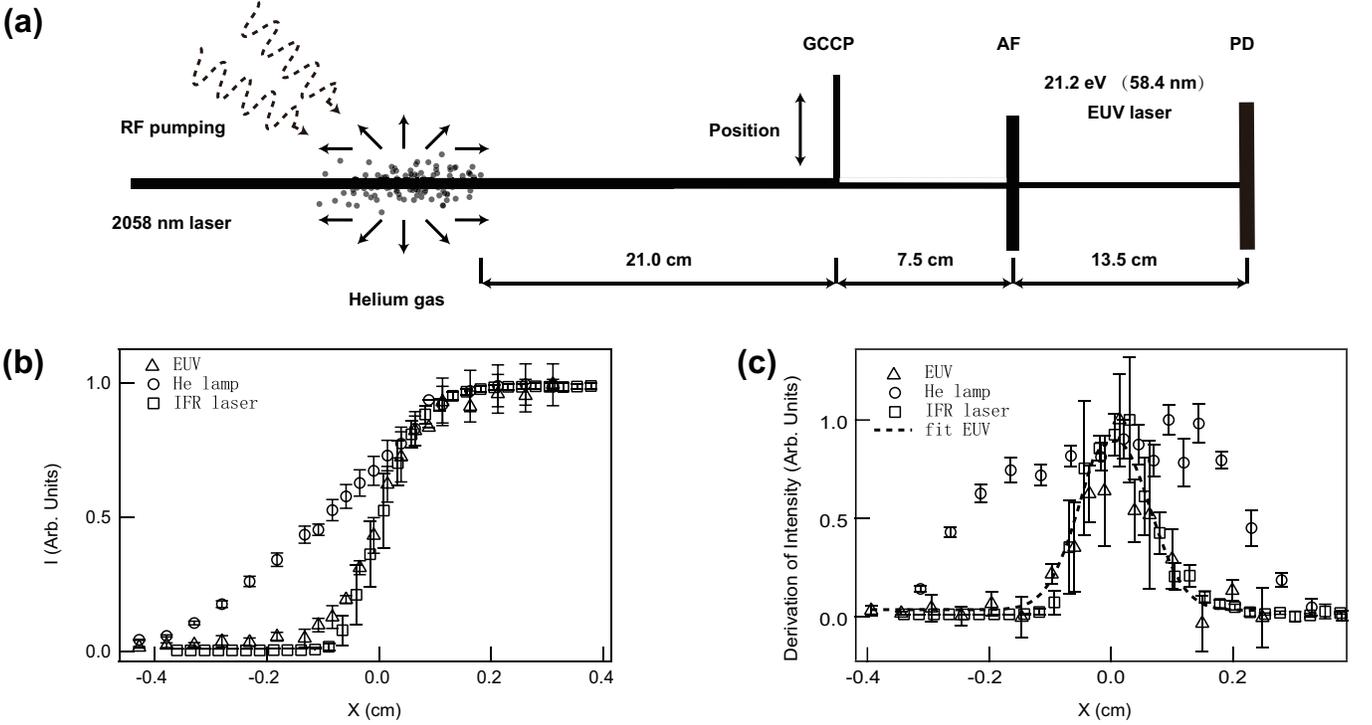}%
		\caption{\label{figs3} Spatial divergence of photon beams. (a) Schematic diagram of the experimental setup including helium lamp, graphite-coated copper plate (GCCP), Aluminum filter (AF), and photodiode (PD). (b) Photocurrents from photodiode irradiated by photons from helium lamp, infrared laser, and the induced EUV laser, respectively, with copper plate at different transverse positions. (c) The derivatives of the intensities in (b).}
	\end{figure*}

	\section{2.3 Spatial coherence of the EUV light}
	We investigated the spatial distribution of the EUV light by knife-edge method\cite{R23} as shown in Fig. 3(a). A graphite-coated copper plate with a straight edge serves as the knife in front of the aluminum filter. The intensity of the beam as shown in Fig. 3(b) was measured by a photodiode and the knife moved transversely to the propagation direction of the laser. The divergence of 2058 nm infra-red laser was measured when the aluminum filter was removed. Then the aluminum filter was moved in, and the helium lamp was turned on to measure the divergence of 58.4 nm spontaneous radiation from the lamp. The photocurrent of induced EUV laser was measured by the difference of photocurrents produced by the photons from excited helium atoms with and without the 2058 nm excitation. The derivatives of photocurrents are shown in Fig. 3(c) and the FWHMs of spatial distribution of 2058 and induced 58.4 nm photons are both 1.3 mm. The distance of 210 mm between the knife and the source allows us to estimate the divergence of the beam. The spot size of the 2058 nm laser at the exit of the microwave cavity is 1.0 mm, and the divergences are estimated to be 1.4 mrad for both 2058 and 58.4 nm lasers. The collimation of induced 58.4 nm photons clearly shows its stimulated character. The derivative of the photocurrent produced by the helium lamp without infra-red excitation is almost flat, indicating its incoherent spontaneous character. Its fast decrease on both sides is caused by the limitations of the photodiode's area (10$\times$10 mm$^2$)

	We also measured the flux of the EUV laser by observing the difference of photocurrents $\Delta$I$_p$ from photodiode behind the aluminum filter with and without excitation laser, which were 2462.0$\pm$2.5 pA and 2449.9$\pm$1.9 pA, respectively. The transmittance of aluminum filter was measured to be 2.73$\%$ (See supplementary Note 1). The flux of the induced EUV laser can be estimated as (5.0$\pm$1.3)$\times$10$^8$ ph/s from $\Delta$I$_p$ of 12.1$\pm$3.1 pA after considering the 0.26 A/W\cite{R24} efficiency of the photodiode.

	The gain $\alpha$ of the stimulated amplification can be estimated from the ratio I$_{st}$/I$_{sp}$ as

	\begin{equation}
	\frac{I_{st}}{I_{sp}}=e^\alpha-1
	\end{equation}

	where I$_{st}$ and I$_{sp}$ are the intensities of stimulated and spontaneous radiation, respectively\cite{R22}. The distribution of induced EUV photons shown in Fig. 3(c) can be fitted by a Gaussian function

	\begin{equation}
	\frac{dI}{dx}=0.025+0.77e^{-(\frac{x}{0.08})^2}
	\end{equation}

	The background and amplitude of the Gaussian function are proportional to the intensities of stimulated and spontaneous photons, respectively. The ratio I$_{st}$/I$_{sp}$ and the gain $\alpha$ are estimated as 31 and 3.5, respectively. The intensity of stimulated radiation is 31 times of magnitude larger than the spontaneous radiation, which confirms the lasing by only a single-pass.

		\begin{figure*}[htbp]
		\includegraphics[width=0.5\textwidth,scale=0.5]{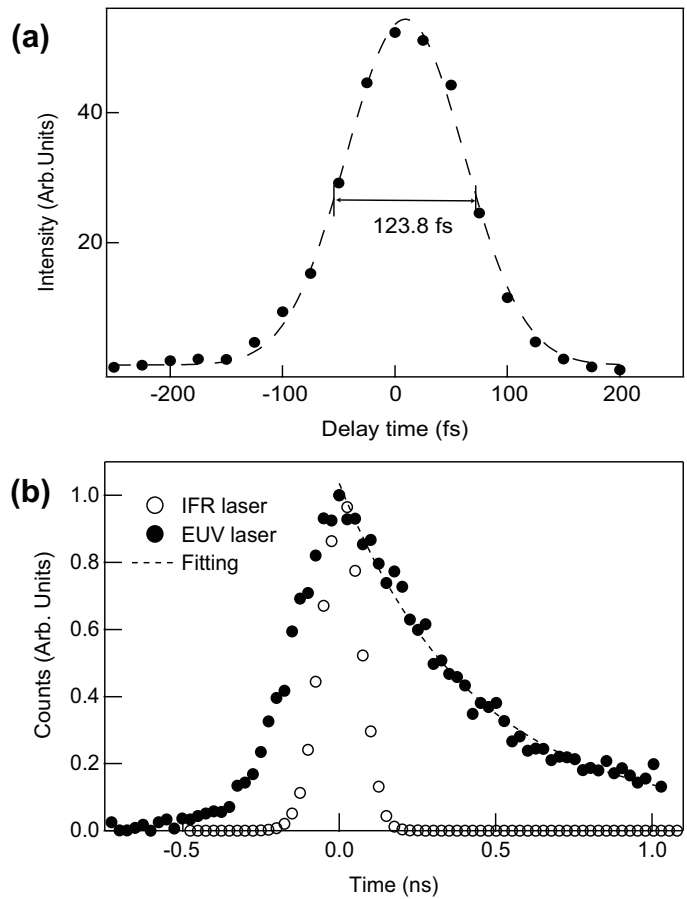}%
		\caption{\label{figs4} Temporal distribution of the infrared and EUV lasers. (a) The intensity of sum-frequency photons with different delay time between 2058 nm and 800 nm lasers. The delay time is controlled by a displacement table with an accuracy of 1$\mu$m (3.3 fs). (b) The temporal distributions of signals from the anode of MCP irradiated by 2058 nm (hollow circles) and 58.4 nm (circles) lasers, respectively.}
	\end{figure*}

	The efficiency of EUV laser observed here is lower than expected mainly due to the low effective power of the infra-red laser. The bandwidth of the infra-red laser is about 49.9 nm, which is too broad compared with the width of the 1s2p level, so only part of the photons has resonant energy, and the effective power can be estimated to be only 23 $\mu$W (See supplementary Note 3). The corresponding effective Rabi oscillation period can be calculated as 9.5 ps, which is much longer than the pulse duration of the excitation laser. Additionally, the too small volume of excited helium atoms by focused microwave is also a possible reason for the weak anti-Stokes laser due to the insufficient atoms at metastable states. However, the power conversion efficiency of this work is 7.4$\times$10$^5$ (See Equation 5), which is 1.1$\times$10$^4$ times and 1.5 times of typical HHG lasers reported\cite{R15,R16}.  In the future, the EUV laser flux can be further improved by using a narrow-band intense infra-red laser and dedicated longer cylindrical excited helium gas\cite{R25}.

	\section{2.4 Temporal profile}

	FThe pulse duration of 2058 nm laser was determined by its cross-correlation with 800 nm laser pulse.  The intensities of sum-frequency photons produced by injecting both 800 nm and 2058 nm lasers into a BBO crystal are measured at different delay time between them as shown in Fig. 4(a) with a correlation of 123.8 fs (FWHM). As the pulse duration of 800 nm laser is 100 fs, the duration of 2058 nm laser can be gotten as 73 fs. Since we cannot perform a similar measurement on the EUV laser, a multi-channel plate (MCP) was used to measure the time profile by recording the time of the signal from MCP anode. The time-resolution of this measurement is limited by the variation of response time of MCP and the electronics, which is about 145 ps according to the observed temporal distribution of signals when the infra-red laser directly impinges on the MCP as shown by hollow circles in Fig. 4(b). The temporal distribution of the EUV photons was measured after filtering the infra-red light with aluminum filter as shown by circles in Fig. 4(b). An FWHM of 426 ps was observed after the 145 ps system broadening was removed from the experimental result of 450 ps. The decay time of 426$\pm$28 ps is obtained by fitting the exponential tail. The decay time is shorter than that of the spontaneous deexcitation from 1s2p to 1s2, which is 556 ps\cite{R26}.  If the lasers are intense enough to make a coherent mixture of the storage and lower levels, the STIRAP can be achieved to produce direct population transfer\cite{R19} between 1$s$2$p$ and 1$s$$^2$ states, and the anti-Stokes laser will have a similar pulse duration as the excitation laser. However, the pulse duration we observed here is too wide compared to that of injected infra-red laser because of the low effective power of the excitation laser and the insufficient coupling between the atoms and lasers discussed above. 
	
	\section{3. Conclusions}
	We successfully developed a practical table top 58.4 nm EUV laser with small divergence and high efficiency based on single-photon related ASRS for the first time. The observed efficiency is higher than that of HHG, and a more intense EUV laser can be expected in the future when narrow band intense infra-red laser and dedicated cylindrical excited atoms are used. The efficiency of the multi-photon process is very low to generate CW light, and the single-photon process reported here can generate CW EUV laser to remedy the space charge effect, which is a bottleneck problem in photoelectron related technology with intense light sources, such as angle-resolved photoemission spectroscopy (ARPES). 
Our success by using single-photon excitation related ASRS opens an applicable way to generate EUV laser with stable and high efficiency compared to nonlinear multi-photon processes. The way we reported here can even generate lasers with shorter wavelengths if other helium-like atoms are used, for example, 19.9 nm and 10.0 nm lasers from Li$^+$ and Be$^{2+}$ ions by injection of 958 and 614 nm excitation lasers, respectively. Our findings confirm a possible path to generate EUV lasers by single-photon related process.

	\section{Acknowledgements}
	This work was supported by the National Key R$\&$D Program of China (Grant No. 2022YFB3608000). S.Q. acknowledges partial support from the National Natural Science Foundation of China (NSFC, Grants No. U1632266, 11927807, and U2032207). Y.W. would like to acknowledge partial support from the Ministry of Science and Technology of China (Grant No. 2017YFA0303000), the NSFC (Grants No. 11827805 and 12150003), and the Shanghai Municipal Science and Technology Major Project (Grant No. 2019SHZDZX01). Z.T. acknowledges partial support from the National Key R$\&$D Program of China (Grant No. 2021YFA1400202), NSFC (Grant No. 11874121), and the Shanghai Municipal Science and Technology Basic Research Project (Grant No. 19JC1410900). Z.L. would like to acknowledge partial support from the NSFC (Grant No. 12222413) and the Natural Science Foundation of Shanghai (Grant No. 22ZR1473300).
We thank Yudong Chen and Zongyuan Fu for the installation of the XUV spectrometer and the code for CCD data analysis, and Denghui Zhang for the code of the Picoammeter. We thank Prof. Chuanshan Tian, Prof. Zhinan Zeng, and Prof. Xiaoyan Liang for fruitful discussions.

\section{Supplementary Note 1}
		\subsection{Materials}
		\textbf{Helium.} Helium with a purity greater than 99.999$\%$ produced by Air Liquide has been used.

		\textbf{Graphite-coated copper Plate (GCCP).} The purity of the copper plate is greater than 99$\%$, and the size is 3$\times$2 cm$^2$ covered by the dag-156 graphite produced by Acheson. 

		\subsection{Methods}
		\textbf{The RF pumping.} The helium lamp was purchased from Fermi instruments in China, and a microwave with a frequency of around 429 MHz was coupled into a quartz tube filled with helium gas to excite helium atoms. The pressure of helium was measured by Inficon's PGE300, and the flow of helium was controlled by a Vacgen's leak valve. Based on the experiments of varying pressure and power (Fig. 11), the optimal experimental conditions were decided as 1 mbar pressure and 60 W RF power. 1 mbar was selected because, with pressure below it, only spectra related to helium II energy levels were observed. An increase in the microwave power intensified both spontaneous and stimulated radiation. When the microwave power was greater than 10 W, the intensity of background spontaneous radiation increased faster and the measurements became difficult when the power was over 60 W. The quartz cavity of the helium lamp is a cylinder with a diameter of 6 mm and a length of 70 mm, and the helium atoms in the cavity are excited by focused microwave.

		\textbf{Measurement of laser spectrum and power.} The wavelength of the infra-red laser was measured by the ideaOptics' NIR2500 spectrometer, and the wavelengths of the 800 nm laser and the sum-frequency photons were measured by OceanOptics' USB2000+ spectrometer. The output power of the 800 nm laser was measured using Coherent's PM30 and PS19 power meters, and that of the 2058 nm laser was measured using Coherent's PS19.
		\textbf{EUV spectrum acquisition.} EUV spectra were observed by h+p's XUV Lightpro spectrometer, with a DO920P-BEN Oxford instruments' Andor CCD camera. The CCD has effective 255$\times$1024 pixels, and the direction along 1024 pixels corresponds to the photon energy (wavelength). In data processing, we integrated all pixels in the direction along 255 pixels to get the intensity at a certain wavelength. Luxel's 120 nm thick aluminum filter (AF) with the nickel-wire support frame is used in front of the XUV Lightpro to block the photons except EUV band. The exposure time of each image was 3 s, and the temperature of CCD was kept at -20 ℃. The dark background with the same exposure time was subtracted.
		\textbf{Experiments with different laser power.} The power of the 2058 nm laser emitted from OPA remained unchanged, and NDC-50C-2 continuous filter purchased from Thorlabs was placed in front of the helium lamp to control the power of the 2058 nm laser injected into the lamp.
		\textbf{Measurements of transmittance of Luxel's 120 nm thick aluminum filter.} A 200 nm thickness aluminum filter (AF) with a diameter of 5 mm was placed between the 120 nm Luxel's AF with a diameter of 10 mm and the helium lamp to block the visible light from the lamp. An AXUV100G photodiode was set behind the 120 nm AF to measure the intensity of photons. The 200 nm AF was set 330 mm away from the lamp, and the Luxel's 120 nm AF was placed 450 mm away from the lamp. This configuration ensures that all the EUV photons through the 200 nm film can also pass through the second one. The Luxel's 120 nm AF was installed on the VAT (01032-CE01-VVA) gate valve, which can be inserted into and extracted off the light path. The photocurrents measured by the photodiode with the Luxel's 120 nm film extract off and insert in the light pass were 1.07 nA and 29.2 pA, respectively, and the transmittance of the Luxel’s 120 nm AF can be estimated to be 2.73$\%$.
		\textbf{Photon flux measurements.} An AXUV100G photodiode from Opto Diode Corporation was used as the detector. The Luxel's 120 nm aluminum filter (AF) with a measured transmittance of 2.73$\%$ is inserted between the helium lamp and the detector. The flux of 58.4 nm laser was measured by the difference of photocurrents measured by the Keithley6485 picoammeter with and without injection of 2058 nm infra-red laser and with the helium lamp turned on. The photocurrents were measured more than 150 times with and without the injected infra-red laser for the evaluation of statistical error.
		\textbf{Temporal profile measurements.} The intensity of sum-frequency 576 nm photons   was measured by OceanOptics' USB2000+ spectrometer, and the temporal distribution of the 58.4 nm laser was measured by the MCP and electronics from Roentdek.
		\section{Supplementary Note 2}
		\textbf{Infra-red Laser generation process.} The seed light is a Ti:Sapphire laser with a wavelength of 800 nm, a pulse duration of 100 fs, a repetition frequency of 10 kHz, and an average power of 10 W. The 800 nm laser was split into two beams with 1:4 intensities. One beam of an average power of 2 W injects into the first BBO crystal to produce white light. Then the white light and another beam of an average power of 8 W hit the second BBO crystal together to generate signal and idler lights by a nonlinear process. The signal and idler of different wavelengths can be obtained by rotating the two BBO crystals. Then both lights pass through two dichroic mirrors, the first one reflects the 800 nm light and transmits the idler and signal band, and the second one reflects the signal band and transmits the idler band. Finally, a laser with a central wavelength of around 2058 nm can be obtained. In another experimental configuration, 800 nm laser and 1348 nm idler hit the BBO crystal together and generate a laser with a central wavelength of 502 nm, and the 800 and 1348 nm photons were moved by a dichroic mirror that only transmits visible light.
		\section{Supplementary Note 3}
		\textbf{Laser power conversion efficiency and Rabi frequency calculations.} Although the power of excitation laser is 0.6 W, the amount of power that can be used effectively is much less. Since the spontaneous radiation lifetime of helium atom 1s2p level is 556 ps, its energy broadening is 5.9×10$^{-7}$ eV due to the uncertainty principle. When using the excitation laser with a central wavelength of 2058 nm and a FWHM bandwidth of 49 nm (Fig. 5), the wavelength range that can be actually used is only 2.0$\times$10$^{-3}$ nm. Because the Gaussian distribution of laser intensity (W/nm) is
		\begin{equation}
		$$I=A$\times$exp(-${\frac{(\lambda-\lambda_c)^2}{2\times\sigma^2}}$)$$
		\end{equation}
	where $\lambda$$_c$ = 2058 nm and $\sigma$=49.9 nm/2.355$\approx$21 nm are the center wavelength and the rms of the distribution.$\int\limits_{-\infty}^{+\infty}Aexp(-{\frac{(\lambda-2058)^2}{2\times21^2}})=\sqrt{2\pi}\times A\times21=0.6$,so A = 0.0115 W/nm
	Therefore, the effective power within a width of 2.0$\times$10$^{-3}$ nm near the Gaussian peak is
	\begin{equation}
	0.0115\times2.0\times10^{-3}=23 \mu W
	\end{equation}

	Therefore, the conversion efficiency can be obtained as
	\begin{equation}
	\eta=\frac{5\times10^8 ph/s\times21.2eV\times1.6\times10^{-19}J/eV}{23\mu W}=7.4times10^{-5}
	\end{equation}

	\newpage
	\begin{figure*}[htbp]
		\includegraphics[width=\linewidth]{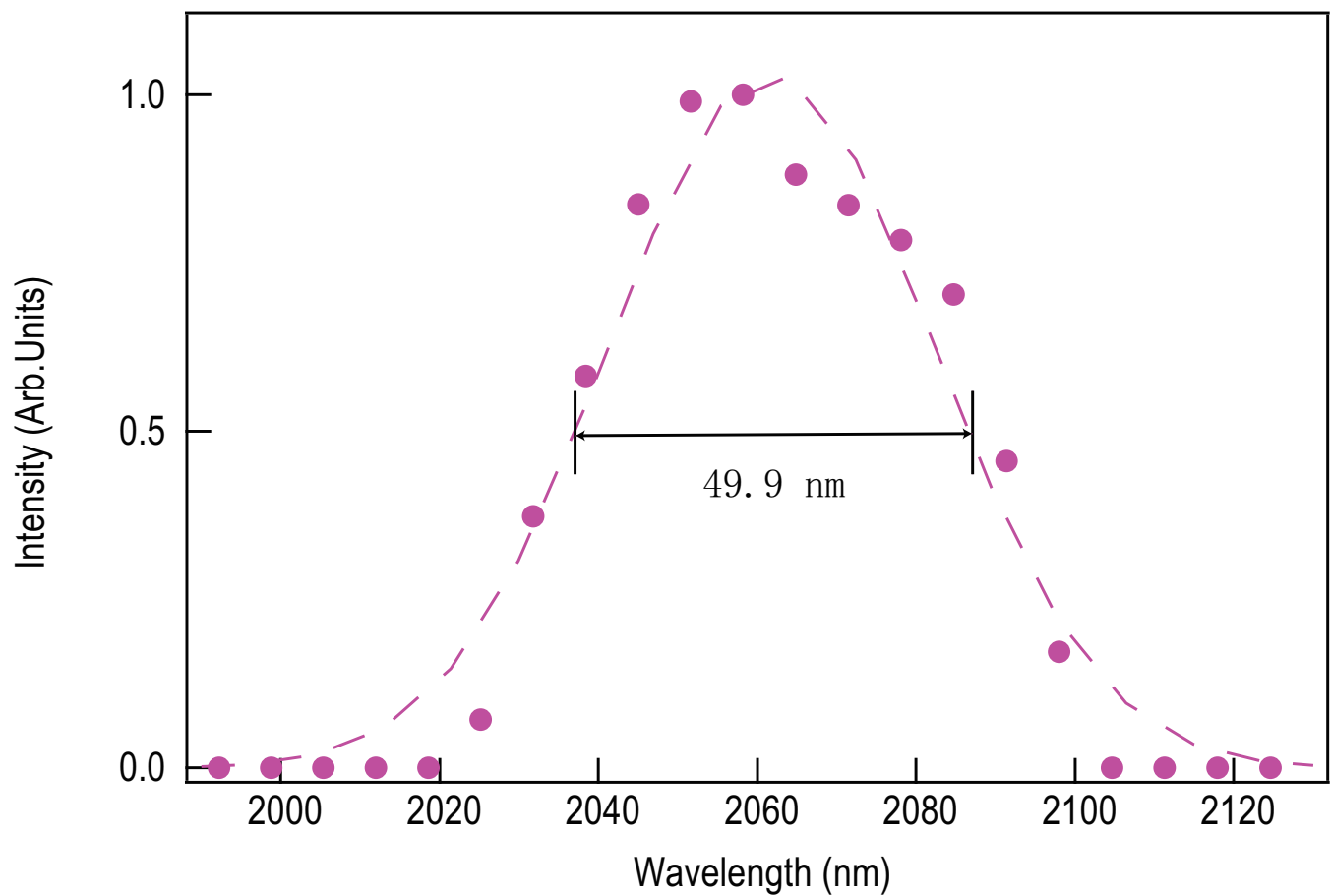}%
		\caption{\label{figs5} Infra-red laser spectrogram with a central wavelength of 2058 nm. The FWHM of Gaussian fitting is 49.9 nm.}
	\end{figure*}

	\begin{figure*}[htbp]
		\includegraphics[width=\linewidth]{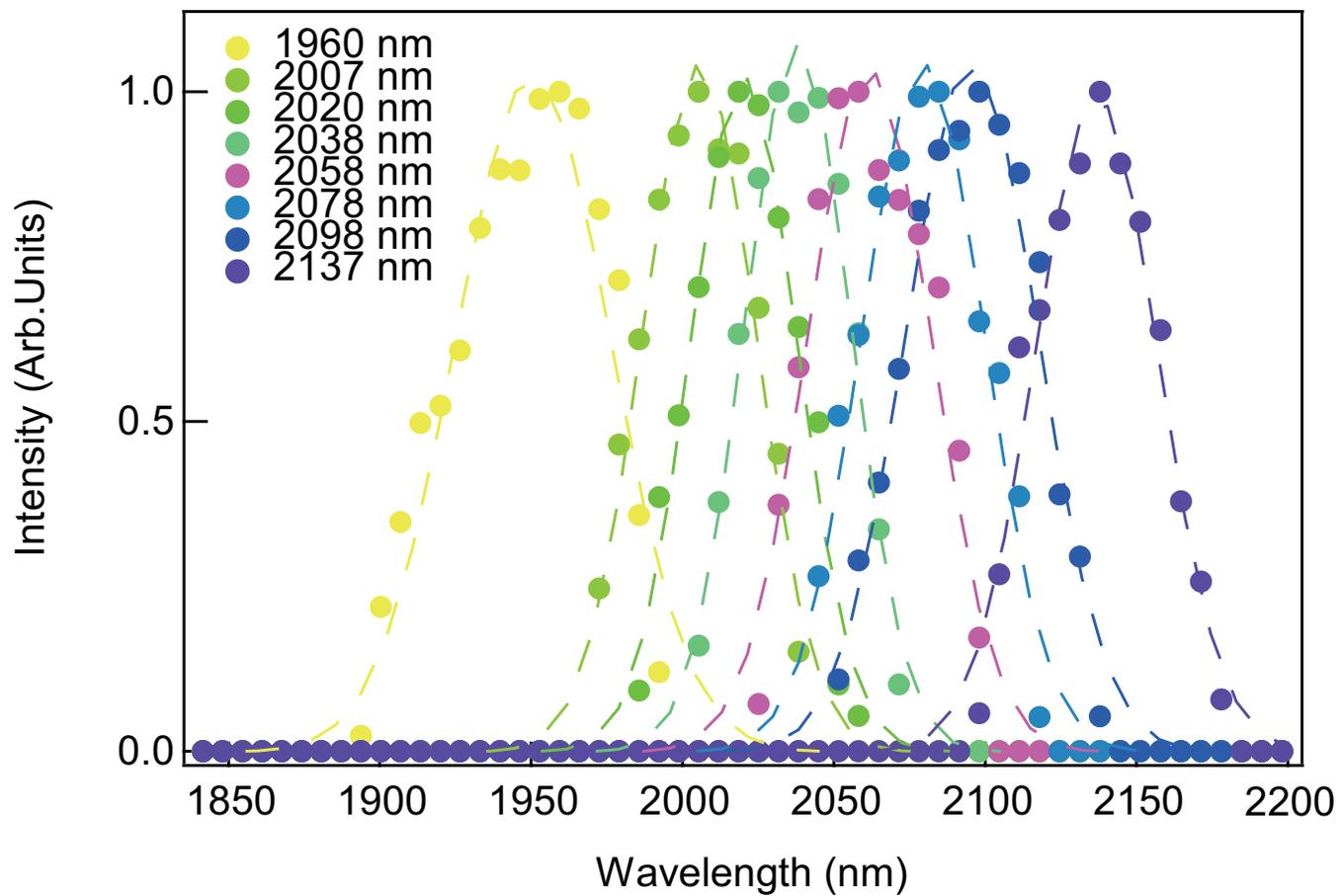}%
		\caption{\label{figs6} Normalized infra-red laser spectrograms of different central wavelengths.}
	\end{figure*}

	\begin{figure*}[htbp]
		\includegraphics[width=\linewidth]{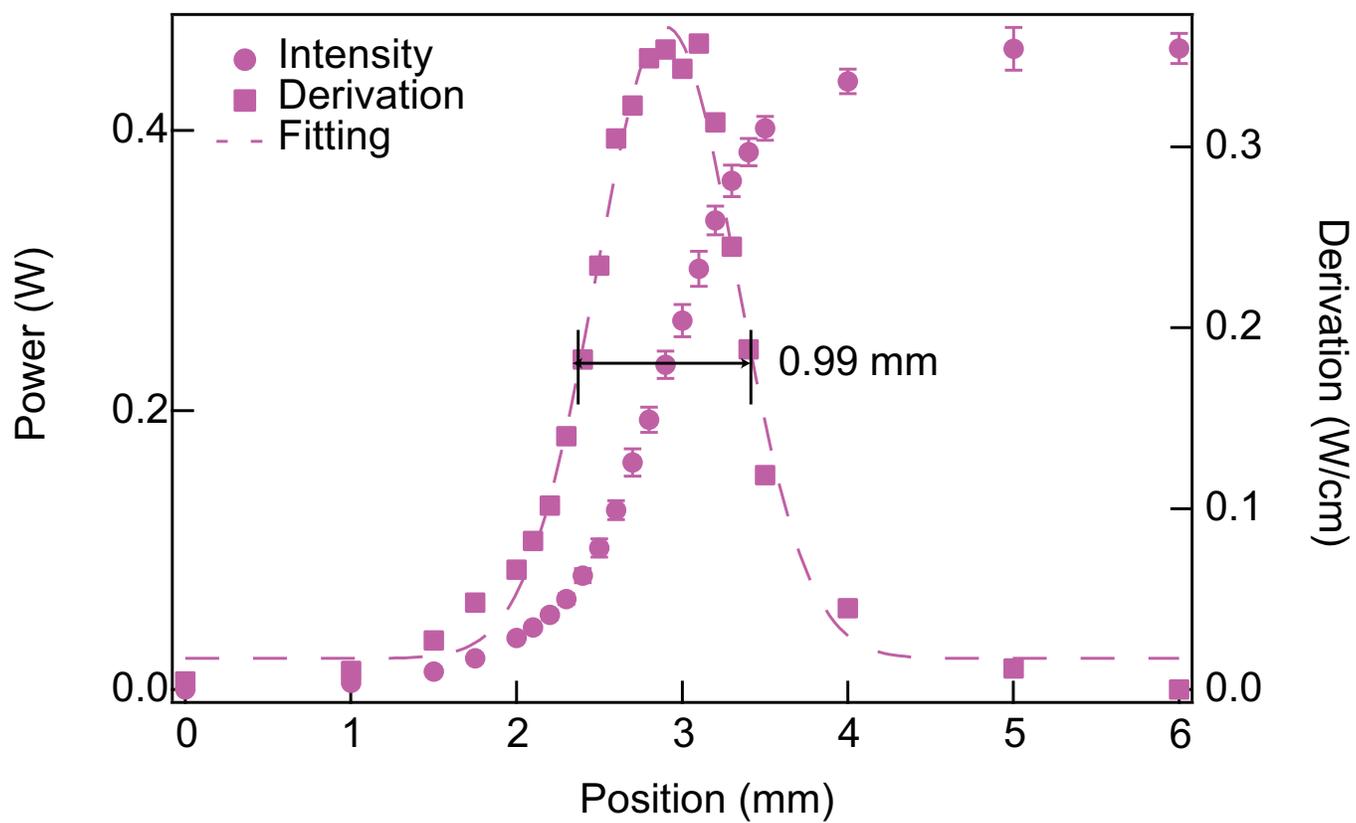}%
		\caption{\label{figs7} The powers of 2058 nm laser measured by power meter with copper plate at different positions at the exit of quartz cavity. The FWHM of infra-red laser spot size is 1.0 mm.}
	\end{figure*}

	\begin{figure*}[htbp]
		\includegraphics[width=\linewidth]{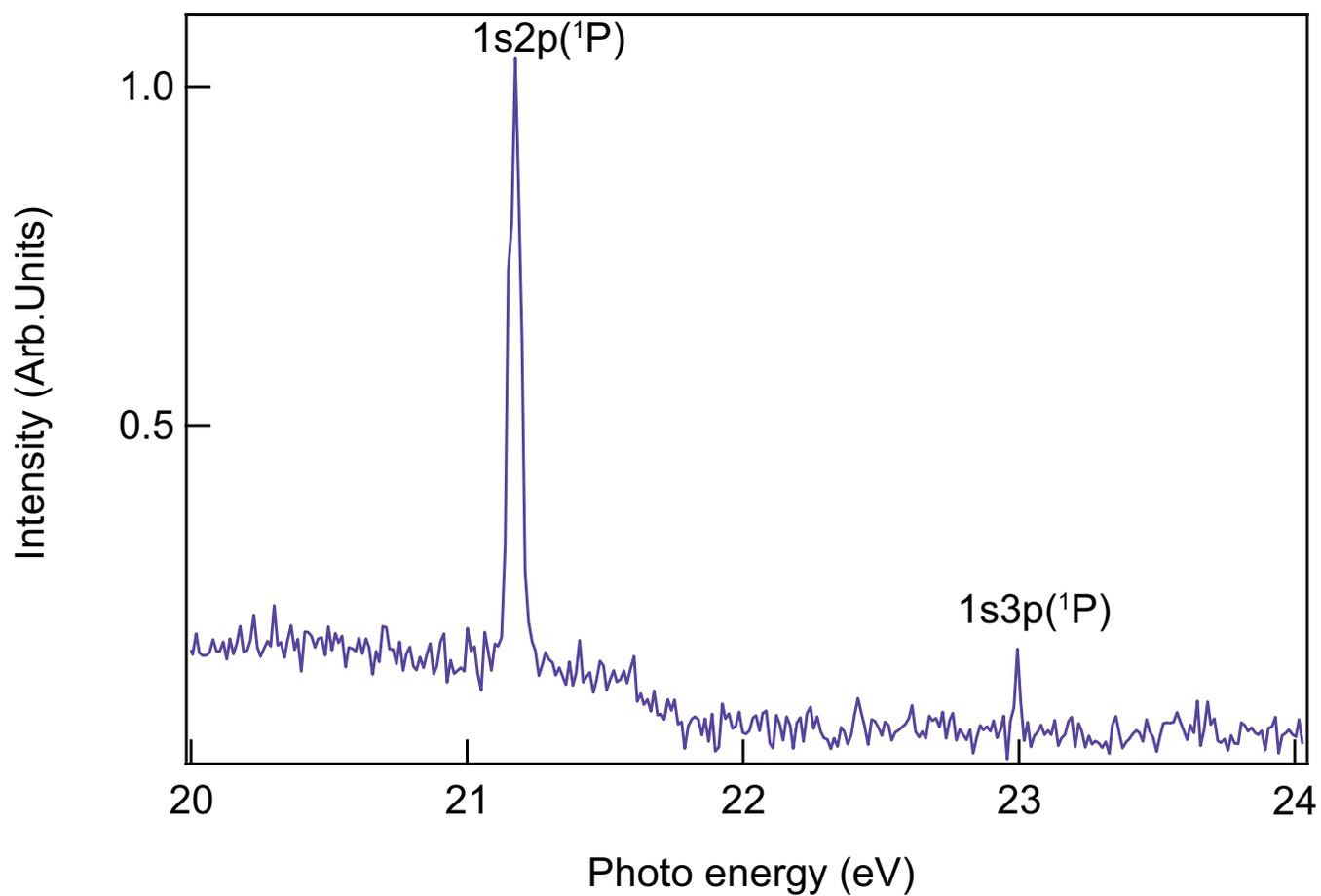}%
		\caption{\label{figs8} The difference of spectra taken by h+p's XUV Lightpro spectrometer with and without excitation laser of a 502 nm central wavelength. The 23.1 eV photons are related to the decay from 1$s$3$p$ to 1$s$$^2$ ground state. Due to the low intensity, its divergence and temporal profile were not studied.}
	\end{figure*}

	\begin{figure*}[htbp]
		\includegraphics[width=\linewidth]{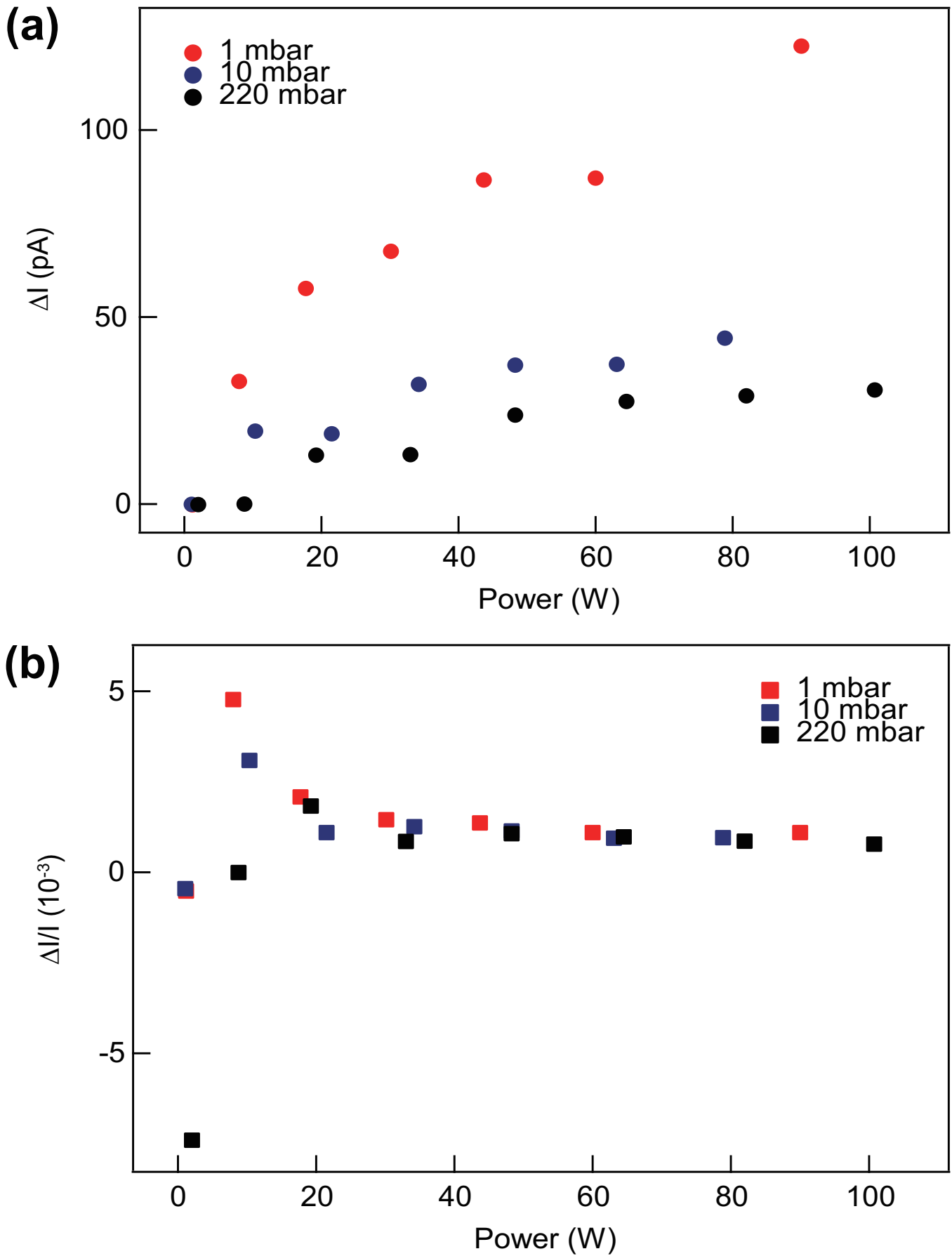}%
		\caption{\label{figs9} Effects of plasma excitation conditions on the intensity of the induced EUV photons. (a) The difference of photocurrents with and without the injection of infra-red laser for different microwave powers and different gas pressures. (b) The intensity ratio of induced photons to the background photons.}
	\end{figure*}

		\section{}
		\subsection{}
		\subsubsection{}

		\bibliography{basename of .bib file}

\begin{thebibliography}{99}
\bibitem{R25} An EUV flux of 1.2$\times$1011 ph/s was obtained by us recently using a dedicated plasma discharge tube and a narrow band 300 mW CW laser.
\end{thebibliography}

\end{document}